\documentclass[aps,twocolumn,groupedaddress]{revtex4}

\usepackage[]{epsfig}

\newcommand {\bra}[1]{{\langle #1 |}}
\newcommand {\ket}[1]{{| #1 \rangle}}

\begin{document}
\bibliographystyle{apsrev}


\title{Spin--Orbit Coupling in the\\ {\it ab initio} Pseudopotential Framework}



\author{G. Theurich}
\email[]{theurich@mrl.ucsb.edu}
\author{N. A. Hill}
\affiliation{Materials Department, University of California, Santa
Barbara, California 93106-5050}


\date{\today}

\begin{abstract}

We describe the implementation of total angular momentum dependent
pseudopotentials in a plane wave formulation of density functional
theory. Our approach thus goes beyond the scalar--relativistic
approximation usually made in {\it ab initio} pseudopotential
calculations and explicitly includes spin--orbit coupling. We
outline the necessary extensions and compare the results to
available all--electron calculations and experimental data.

\end{abstract}

\maketitle



%
%

The {\it ab initio} pseudopotential method \cite{Ihm79p4409,
Payne92p1045, Kresse96p15} has become a standard tool in many
areas of electronic structure calculation. Even magnetic compounds
containing $3d$ transition metal ions lie in the realm of the
plane wave pseudopotential approach of density functional theory
\cite{Sasaki95p12760, Moroni97p15629}. In order to obtain high
precision results it is necessary to include relativistic effects
when calculating the electronic structure of materials containing
third row elements \cite{Bachelet82p2103}. Hence it is now
standard procedure to create scalar--relativistic pseudopotentials
that include the kinematic relativistic effects (mass--velocity
and Darwin term) from the fully relativistic all--electron
solution of the atom \cite{Bachelet82p4199, Hamann89p2980,
Rappe90p1227, Troullier91p1993, Vanderbilt90p7892}.

The spin--orbit interaction, however, is only effectively taken
into account by the construction of $j$--averaged pseudopotentials
for each angular momentum $l$. Thus no spin--orbit splitting is
present in the resulting band structure. Although the
scalar--relativistic approximation is acceptable in many
situations it becomes insufficient in cases where the observed
quantities, such as hole effective masses or spin relaxation times,
are a direct consequence of the  spin--orbit splitting
\cite{Cardona88p1806}.

In this paper we report on the implementation of spin--orbit
coupling in the pseudopotential scheme. We give the equations
required to program the formalism within a generalized spinor
approach, and compare the results to experimental data and to
all-electron calculations. This is, to our knowledge, the first
zeroth order implementation of spin--orbit coupling in the {\it ab
initio} pseudopotential scheme. Prior publications on this matter
always relied on a second variation of the scalar--relativistic
zeroth order eigenstates, including spin--orbit coupling to first
order in perturbation theory \cite{Hybertsen86p2920, Surh91p4286,
Hemstreet93p4238}.

Although the fully relativistic treatment of the problem would
require a four--current formulation with Dirac spinors it has been
shown by Kleinman that a Pauli--like Schr\"odinger equation
captures all relativistic effects to order $\alpha^2$, where
$\alpha$ is the fine structure constant \cite{Kleinman80p2630}.
The total ionic pseudopotential to be used is
\begin{equation}
\label{j_psp}
V_{PS} =\sum_{l,j,m_j}\ket{\Phi^{l,j}_{m_j}}V_{l,j}\bra{\Phi^{l,j}_{m_j}}\ ,
\end{equation}
where the $\ket{\Phi^{l,j}_{m_j}}$ are the total angular momentum
eigenfunctions which can be written in terms of the spherical
harmonics, $Y^m_l$, and the eigenfunctions of the $z$--component
of the Pauli spin operator, $\ket{\uparrow}$ and $\ket{\downarrow}$.
For $j=l+\frac{1}{2}$, $m_j=m+\frac{1}{2}$ the $\ket{\Phi^{l,j}_{m_j}}$ equal
\begin{equation} \left(\frac{l+m+1}{2l+1}\right)^\frac{1}{2}\ket{Y^m_l}\ket{\uparrow}+         \left(\frac{l-m}{2l+1}\right)^\frac{1}{2}\ket{Y^{m+1}_l}\ket{\downarrow}
\end{equation}
and for $j=l-\frac{1}{2}$, $m_j=m-\frac{1}{2}$ have the form
\begin{equation} \left(\frac{l-m+1}{2l+1}\right)^\frac{1}{2}\ket{Y^{m-1}_l}\ket{\uparrow}-        \left(\frac{l+m}{2l+1}\right)^\frac{1}{2}\ket{Y^m_l}\ket{\downarrow}
\label{j_minus}\ .
\end{equation}
Hence the operator $V_{PS}$ acts in
both orbital and spin space. Note that there is only one radial
pseudopotential component $V_{l,j}$ with $j=\frac{1}{2}$ for $l=0$
but two with $j=l+\frac{1}{2}$ and $j=l-\frac{1}{2}$ for each
$l>0$. The index $m_j$ in equation (\ref{j_psp}) runs from $-j$ to
$+j$. It is computationally more efficient to transcribe each term
of the semi--local pseudopotential operator $V_{PS}$ into the
fully separable Kleinman--Bylander (KB) form
\cite{Kleinman82p1425}
\begin{equation}
\label{relKB} V_{KB} = \sum_{i_s,
i_a}\sum_{l,j,m_j}\frac{\ket{\delta
V^{i_s,i_a}_{l,j}\phi^{i_s,i_a}_{l,j,m_j}}\bra{\phi^{i_s,i_a}_{l,j,m_j}\delta
V^{i_s,i_a}_{l,j}}}{\bra{\phi^{i_s}_{l,j,m_j}}\delta
V^{i_s}_{l,j}\ket{\phi^{i_s}_{l,j,m_j}}}
\end{equation}
using the solutions of the atomistic pseudopotential problem
\begin{equation}
\ket{\phi^{i_s,i_a}_{l,j,m_j}}=\ket{R^{i_s,i_a}_{l,j}}\ket{\Phi^{l,j}_{m_j}}\ ,
\end{equation}
where $\ket{R^{i_s,i_a}_{l,j}}$ is the radial part of the pseudo
eigenfunction of atom species $i_s$ at position $r_{i_s,i_a}$. The
potential $\delta V^{i_s,i_a}_{l,j}$ is defined as the difference
\begin{equation}
\label{minus_local}
\delta V^{i_s,i_a}_{l,j}(r) = V_{l,j}(r-r_{i_s,i_a}) - V_{loc}(r-r_{i_s,i_a})\ ,
\end{equation}
where $V_{loc}(r)$ is an arbitrary local potential that needs to
be chosen such that the remaining $\delta V$'s are short ranged.
The complete KB pseudopotential operator is thus given as the sum
of the local part and the non--local KB operator.

To our knowledge all previous pseudopotential calculations that
included spin--orbit coupling did so by using a second variation
step on the scalar--relativistic zeroth order wave functions, thus
including the spin--orbit term to first order perturbation theory
\cite{Hybertsen86p2920, Surh91p4286, Hemstreet93p4238}. In
contrast we solve directly for general two--component spinor Bloch
wave functions expanding in a plane wave spinor basis
\begin{equation}
\label{plane_wave}
\ket{\psi_{nk}} = \sum_{G,\sigma}c^{n,k}_{G,\sigma}\ket{k+G}\ket{\sigma},
\end{equation}
where $G$ are reciprocal lattice vectors and $\sigma$ sums over up
and down spin. In the basis of equation (\ref{plane_wave}) the action of the KB
operator is as follows:
\begin{widetext}
\begin{equation}
\label{KBelement}
\bra{\sigma}\bra{k+G}V_{KB}\ket{\psi_{nk}}=\sum_{i_s, i_a}\sum_{l,j,m_j}D^{KB}_{i_s,l,j}\ \varphi^{i_s,i_a}_{k+G}\
M^{KB, \sigma}_{i_s,l,j,m_j,k+G}\ f^{KB}_{i_s,i_a,k,n,l,j,m_j}\ ,
\end{equation}
\end{widetext}
where
\begin{equation}
D^{KB}_{i_s,l,j} = \left(\frac{4\pi}{V}\right)\frac{1}{\int dr\ r^2 R^{*,i_s}_{l,j}(r)\ \delta V^{i_s}_{l,j}(r)\ R^{i_s}_{l,j}(r)}
\end{equation}
and
\begin{equation}
\varphi^{i_s,i_a}_{k+G} = e^{-i(\vec k+\vec G)\cdot {\vec r}_{i_s,
i_a}}
\end{equation}
is a phase factor associated with the atomic position. The spin
dependent factor $M^{KB}$ of equation (\ref{KBelement}) can be written as a spinor and for
$j=l+\frac{1}{2}$ is
\begin{equation}
\label{M_plus}
M^{KB, \sigma}_{i_s,l,j,m_j,k+G} =
{
\sqrt{l+m+1}\ F^{KB}_{i_s,l,j,m,k+G}
\choose
\sqrt{l-m}\ F^{KB}_{i_s,l,j,m+1,k+G}
}
\end{equation}
and
\begin{equation}
\label{M_minus}
M^{KB, \sigma}_{i_s,l,j,m_j,k+G} =
{
\sqrt{l-m+1}\ F^{KB}_{i_s,l,j,m-1,k+G}
\choose
-\sqrt{l+m}\ F^{KB}_{i_s,l,j,m,k+G}
}
\end{equation}
for $j=l-\frac{1}{2}$. Also the last factor of equation (\ref{KBelement}),
$f^{KB}_{i_s,i_a,k,n,l,j,m_j}$, depends on $j$ as follows
\begin{eqnarray}
&\sum_{G'} \varphi^{* i_s,i_a}_{k+G}&
(c^{n,k}_{G',\uparrow}\sqrt{l+m+1}\ F^{* KB}_{i_s,l,j,m,k+G'}\nonumber\\
&&+c^{n,k}_{G',\downarrow}\sqrt{l-m}\ F^{* KB}_{i_s,l,j,m+1,k+G'})
\end{eqnarray}
for $j=l+\frac{1}{2}$ and
\begin{eqnarray}
&\sum_{G'} \varphi^{* i_s,i_a}_{k+G}&
(c^{n,k}_{G',\uparrow}\sqrt{l-m+1}\ F^{* KB}_{i_s,l,j,m-1,k+G'}\nonumber\\
&&-c^{n,k}_{G',\downarrow}\sqrt{l+m}\ F^{* KB}_{i_s,l,j,m+1,k+G'})
\end{eqnarray}
for $j=l-\frac{1}{2}$. Finally the KB factors $F^{KB}$ appearing
in equations (\ref{M_plus}) and (\ref{M_minus}) are defined as
\begin{eqnarray}
F^{KB}_{i_s,l,j,m,k+G} =\sqrt{\frac{4\pi}{2l+1}}Y^m_l(\theta, \varphi)\times\nonumber\\
\int dr\ r^2\ j_l(|k+G|r)\ \delta V^{i_s}_{l,j}(r)\ R^{i_s}_{l,j}(r),
\end{eqnarray}
where $Y^m_l$ are the spherical harmonics, the polar angles
$\theta$ and $\varphi$ are determined by the vector $\vec k + \vec G$
and $j_l$ are the spherical Bessel functions. The KB factors are
calculated once and stored in memory. The contribution of a state
$\ket{\psi_{nk}}$ to the non--local KB part of the total energy is
thus given by
\begin{equation}
E^{KB}=\sum_{i_s, i_a}\sum_{l,j,m_j}D^{KB}_{i_s,l,j}\left|
f^{KB}_{i_s,i_a,k,n,l,j,m_j}
\right|^2\ .
\end{equation}

In order to test our implementation we calculated the properties
of GaAs and compared with all--electron calculations and available
experimental data. Self--consistency was achieved by direct
minimization of the total energy via a conjugate gradient method
\cite{Payne92p1045}. The gallium and arsenic pseudopotentials were
created following the Troullier--Martins scheme
\cite{Troullier91p1993}, and both contained $s$ and $p$
components. Care must be given to the local part of the
pseudopotential entering in equation (\ref{minus_local}) to ensure
good transferability. We used the $j$--average of the unbound $4d$
state in case of gallium and likewise the $j$--average of the $p$
states for the local part of the arsenic pseudopotential.

\begin{table}
\begin{tabular}{l|cc|c|c}
& S-FKKR$^a$ & S-FLAPW$^b$ & R-PWPP & Exp.$^c$\\
\hline
$a_0$ (\AA) & 5.56 & 5.620 & 5.642 & 5.653\\
$B_0$ (GPa) & 77 & 74 & 72.2 & 74.8\\
\end{tabular}
\caption{Equilibrium lattice constant and bulk modulus determined
in this work (R-PWPP) compared to all--electron calculations $^a$
Scalar relativistic FKKR, M. Asato et al., PRB {\bf 60}, 5202
(1999), $^b$ Scalar relativistic FLAPW, C. Filippi et al., PRB
{\bf 50}, 14947 (1994) and experiment $^c$ Landolt--B\"ornstein,
Vol 22 (1987)} \label{tab_structural}
\end{table}

\begin{figure}[bp]
\epsfig{file=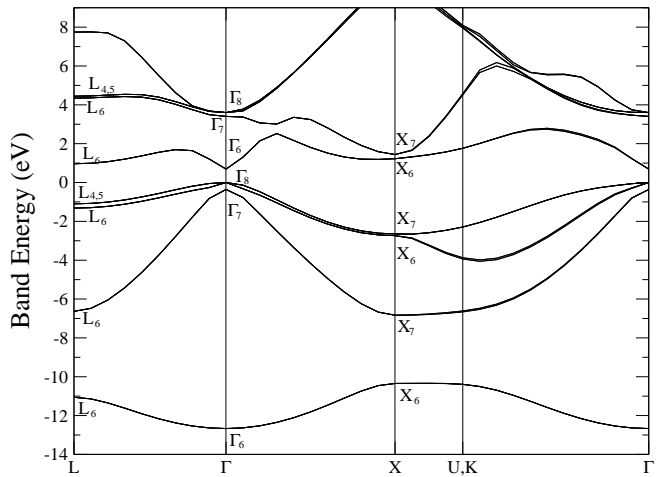, width=\columnwidth, angle=0} \caption{Fully
relativistic bandstructure of GaAs obtained with the implementation
described in this work.} \label{fig_band}
\end{figure}

The results for lattice constant and bulk modulus for GaAs are
shown in table \ref{tab_structural}. The good agreement between
our fully relativistic pseudopotential results and the
scalar--relativistic all--electron values in table
\ref{tab_structural} confirms our approach and reaffirms the
notion that spin--orbit splittings have little effect on the
structural properties of semiconductors \cite{Bachelet85p879}. Our
calculated bandstructure in figure \ref{fig_band} on the other
hand shows clear evidence of spin--orbit coupling. The top of the
valence band splits into the light hole, heavy hole manifold and
the split off band, separated by 350meV. A similar split is also
observed in the upper conduction bands at the Brillouin zone
center. In table \ref{tab_splitting} we compare the characteristic
spin--orbit splittings we obtained for GaAs at the experimental
lattice constant with values from two fully relativistic
all--electron calculations found in the literature. The agreement
with both all--electron calculations is excellent.

\begin{table}[h]
\begin{tabular}{l|rr|r}
Splitting & R-FKKR$^a$ & R-FLAPW$^b$ & R-PWPP\\
\hline
$\Delta_0(\Gamma_{15}^v)$ & 0.35 & 0.34 & 0.35\\
$\Delta'_0(\Gamma_{15}^c)$ & 0.20 & -- & 0.20\\
$\Delta_1(L_{3}^v)$ & 0.09 & 0.09 & 0.09\\
$\Delta(X_{5}^v)$ & 0.21 & 0.20 & 0.22\\
\end{tabular}
\caption{Spin--orbit splittings for GaAs obtained in this work
(R-PWPP) compared to the results of two relativistic all--electron
calculations: $^a$ Fully relativistic FKKR, S. Bei der Kellen, A.
J. Freeman, PRB {\bf 54}, 11187 (1996) and $^b$ Scalar
relativistic FLAPW + 2nd variation, C. Filippi, et al., PRB {\bf
50}, 14947 (1994)} \label{tab_splitting}
\end{table}

For completeness we compare in table \ref{tab_eigenvalues} the
eigenvalue spectrum at three special $k$ points of the Brillouin
zone with the eigenvalues obtained from the two fully relativistic
all--electron calculations cited in table \ref{tab_splitting}. The
zero of energy was chosen to coincide with the top of the valence
band. Despite the generally good agreement there are two obvious
discrepancies at the Brillouin zone center that need some
clarification. First the direct band gap of our pseudopotential
calculation at $\Gamma$ is more than 4 times larger than the gap
resulting from the all--electron calculations. Second the valence
band width of our approach is slightly smaller compared to the
all--electron results. Both discrepancies result from the fact
that the gallium $3d$ orbitals were placed in the frozen core in
our calculation but are free to change in the all--electron
approaches. Due to the well known self--interaction problem of the
local density approximation to density functional theory
\cite{Perdew81p5048} these fairly localized states will lie too
high in energy when not frozen. The symmetry of the $d$ states in
the zincblende lattice at $\Gamma$ only allows hybridization with
$p$ states, e.g. the top of the valence band. Hence the top of the
valence band will shift upwards, leading to a reduced band gap and
at the same time an increase in the valence band width. Due to the
mixed character of the band states away from the Brillouin zone
center the effect of the gallium $3d$ states is most pronounced at
$\Gamma$.

\begin{table}
\begin{tabular}{c|rr|r}
Level & R-FKKR$^a$ & R-FLAPW$^b$ & R-PWPP\\
\hline
$\Gamma_6^v$ & -12.94 & -12.91 & -12.67\\
$\Gamma_7^v$ & -0.35 & -0.34 & -0.35\\
$\Gamma_8^v$ & 0.00 & 0.00 & 0.00\\
$\Gamma_6^c$ & 0.12 & 0.17 & 0.69\\
$\Gamma_7^c$ & 3.46 & & 3.40\\
$\Gamma_8^c$ & 3.66 & & 3.60\\
\newline\\
$X_6^v$ & -10.42 & -10.41 & -10.35\\
$X_7^v$ & -7.02 & -7.00 & -6.83\\
$X_6^v$ & -2.88 & -2.85 & -2.74\\
$X_7^v$ & -2.79 & -2.76 & -2.65\\
$X_6^c$ & 1.17 & 1.23 & 1.23\\
$X_7^c$ & 1.39 & & 1.44\\
\newline\\
$L_6^v$ & -11.18 & -11.14 & -11.06\\
$L_6^v$ & -6.83 & -6.82 & -6.63\\
$L_6^v$ & -1.38 & -1.36 & -1.32\\
$L_{4,5}^v$ & -1.17 & -1.16 & -1.10\\
$L_6^c$ & 0.71 & 0.73 & 0.97\\
$L_6^c$ & 4.38 & & 4.34\\
$L_{4,5}^c$ & 4.46 & & 4.44\\
\end{tabular}
\caption{Eigenvalue spectrum at three special $k$ points compared
with the same relativistic all--electron calculations cited in
table \ref{tab_splitting}} \label{tab_eigenvalues}
\end{table}

Compared to calculations that do not include the spin--orbit term
we find that the inclusion of spin--orbit coupling worsens the
short coming of the local density approximation of underestimating
the band gap. The reason for this observation simply lies in the
fact that the top of the valence band splits and the light and
heavy hole states move closer to the bottom of the conduction
band.

In conclusion, we have implemented spin--orbit coupling in the
well established {\it ab initio} pseudopotential approach of
density functional theory. This paper gives the necessary
expressions in a two--component spinor plane wave basis and
demonstrates the applicability of the method for bulk GaAs. Our
results compare very well to relativistic all--electron
calculations. Since our direct approach is based on a complete
spinor plane wave basis it can easily be extended to systems that
show exchange splitting and exhibit non--collinear spin
arrangements.

The code will be available under the GNU General Public License
\cite{GPL} at \url{http://www.mrl.ucsb.edu/~theurich/Spinor/}.

This work was supported by the ONR grant number N00014-00-10557,
by NSF-DMR under the grant 9973076 and by ACS PRF under the grant
33851-G5. G.T. acknowledges fellowship support from the UCSB
Materials Research Lab., funded by the Corning Foundation.

%


\bibliography{../../PhDThesis/MyBib/mybib}

\begin{thebibliography}{10}
\expandafter\ifx\csname bibnamefont\endcsname\relax
  \def\bibnamefont#1{#1}\fi
\expandafter\ifx\csname bibfnamefont\endcsname\relax
  \def\bibfnamefont#1{#1}\fi
\expandafter\ifx\csname url\endcsname\relax
  \def\url#1{\texttt{#1}}\fi
\expandafter\ifx\csname urlprefix\endcsname\relax\def\urlprefix{URL }\fi
\providecommand{\bibinfo}[2]{#2}
\providecommand{\eprint}[2][]{\url{#2}}

\bibitem{Ihm79p4409}
\bibinfo{author}{\bibfnamefont{J.}~\bibnamefont{Ihm}},
  \bibinfo{author}{\bibfnamefont{A.}~\bibnamefont{Zunger}}, \bibnamefont{and}
  \bibinfo{author}{\bibfnamefont{M.~L.} \bibnamefont{Cohen}},
  \bibinfo{journal}{J. Phys. C: Solid State Phys.}
  \textbf{\bibinfo{volume}{{\bf 12}}}, \bibinfo{pages}{\ 4409}
  (\bibinfo{year}{1979}).

\bibitem{Payne92p1045}
\bibinfo{author}{\bibfnamefont{M.~C.} \bibnamefont{Payne}},
  \bibinfo{author}{\bibfnamefont{M.~P.} \bibnamefont{Teter}},
  \bibinfo{author}{\bibfnamefont{D.~C.} \bibnamefont{Allan}},
  \bibinfo{author}{\bibfnamefont{T.~A.} \bibnamefont{Arias}}, \bibnamefont{and}
  \bibinfo{author}{\bibfnamefont{J.~D.} \bibnamefont{Joannopoulos}},
  \bibinfo{journal}{Rev. Mod. Phys.} \textbf{\bibinfo{volume}{{\bf 64}}},
  \bibinfo{pages}{\ 1045} (\bibinfo{year}{1992}).

\bibitem{Kresse96p15}
\bibinfo{author}{\bibfnamefont{G.}~\bibnamefont{Kresse}} \bibnamefont{and}
  \bibinfo{author}{\bibfnamefont{J.}~\bibnamefont{Furthm{\"u}ller}},
  \bibinfo{journal}{Comp. Mat. Sci.} \textbf{\bibinfo{volume}{{\bf 6}}},
  \bibinfo{pages}{\ 15} (\bibinfo{year}{1996}).

\bibitem{Sasaki95p12760}
\bibinfo{author}{\bibfnamefont{T.}~\bibnamefont{Sasaki}},
  \bibinfo{author}{\bibfnamefont{A.~M.} \bibnamefont{Rappe}}, \bibnamefont{and}
  \bibinfo{author}{\bibfnamefont{S.~G.} \bibnamefont{Louie}},
  \bibinfo{journal}{Phys. Rev. B} \textbf{\bibinfo{volume}{{\bf 52}}},
  \bibinfo{pages}{\ 12760} (\bibinfo{year}{1995}).

\bibitem{Moroni97p15629}
\bibinfo{author}{\bibfnamefont{E.~G.} \bibnamefont{Moroni}},
  \bibinfo{author}{\bibfnamefont{G.}~\bibnamefont{Kresse}},
  \bibinfo{author}{\bibfnamefont{J.}~\bibnamefont{Hafner}}, \bibnamefont{and}
  \bibinfo{author}{\bibfnamefont{J.}~\bibnamefont{Furthm{\"u}ller}},
  \bibinfo{journal}{Phys. Rev. B} \textbf{\bibinfo{volume}{{\bf 56}}},
  \bibinfo{pages}{\ 15629} (\bibinfo{year}{1997}).

\bibitem{Bachelet82p2103}
\bibinfo{author}{\bibfnamefont{G.~B.} \bibnamefont{Bachelet}} \bibnamefont{and}
  \bibinfo{author}{\bibfnamefont{M.}~\bibnamefont{Schl{\"u}ter}},
  \bibinfo{journal}{Phys. Rev. B} \textbf{\bibinfo{volume}{{\bf 25}}},
  \bibinfo{pages}{\ 2103} (\bibinfo{year}{1982}).

\bibitem{Bachelet82p4199}
\bibinfo{author}{\bibfnamefont{G.~B.} \bibnamefont{Bachelet}},
  \bibinfo{author}{\bibfnamefont{D.~R.} \bibnamefont{Hamann}},
  \bibnamefont{and}
  \bibinfo{author}{\bibfnamefont{M.}~\bibnamefont{Schl{\"u}ter}},
  \bibinfo{journal}{Phys. Rev. B} \textbf{\bibinfo{volume}{{\bf 26}}},
  \bibinfo{pages}{\ 4199} (\bibinfo{year}{1982}).

\bibitem{Hamann89p2980}
\bibinfo{author}{\bibfnamefont{D.~R.} \bibnamefont{Hamann}},
  \bibinfo{journal}{Phys. Rev. B} \textbf{\bibinfo{volume}{{\bf 40}}},
  \bibinfo{pages}{\ 2980} (\bibinfo{year}{1989}).

\bibitem{Rappe90p1227}
\bibinfo{author}{\bibfnamefont{A.~M.} \bibnamefont{Rappe}},
  \bibinfo{author}{\bibfnamefont{K.~M.} \bibnamefont{Rabe}},
  \bibinfo{author}{\bibfnamefont{E.}~\bibnamefont{Kaxiras}}, \bibnamefont{and}
  \bibinfo{author}{\bibfnamefont{J.~D.} \bibnamefont{Joannopoulos}},
  \bibinfo{journal}{Phys. Rev. B} \textbf{\bibinfo{volume}{{\bf 41}}},
  \bibinfo{pages}{\ 1227} (\bibinfo{year}{1990}).

\bibitem{Troullier91p1993}
\bibinfo{author}{\bibfnamefont{N.}~\bibnamefont{Troullier}} \bibnamefont{and}
  \bibinfo{author}{\bibfnamefont{J.~L.} \bibnamefont{Martins}},
  \bibinfo{journal}{Phys. Rev. B} \textbf{\bibinfo{volume}{{\bf 43}}},
  \bibinfo{pages}{\ 1993} (\bibinfo{year}{1991}).

\bibitem{Vanderbilt90p7892}
\bibinfo{author}{\bibfnamefont{D.}~\bibnamefont{Vanderbilt}},
  \bibinfo{journal}{Phys. Rev. B} \textbf{\bibinfo{volume}{{\bf 41}}},
  \bibinfo{pages}{\ 1990} (\bibinfo{year}{1990}).

\bibitem{Cardona88p1806}
\bibinfo{author}{\bibfnamefont{M.}~\bibnamefont{Cardona}},
  \bibinfo{author}{\bibfnamefont{M.~E.} \bibnamefont{Christensen}},
  \bibnamefont{and} \bibinfo{author}{\bibfnamefont{G.}~\bibnamefont{Fasol}},
  \bibinfo{journal}{Phys. Rev. B} \textbf{\bibinfo{volume}{{\bf 38}}},
  \bibinfo{pages}{\ 1806} (\bibinfo{year}{1988}).

\bibitem{Surh91p4286}
\bibinfo{author}{\bibfnamefont{M.~P.} \bibnamefont{Surh}},
  \bibinfo{author}{\bibfnamefont{M.-F.} \bibnamefont{Li}}, \bibnamefont{and}
  \bibinfo{author}{\bibfnamefont{S.~G.} \bibnamefont{Louie}},
  \bibinfo{journal}{Phys. Rev. B} \textbf{\bibinfo{volume}{{\bf 43}}},
  \bibinfo{pages}{\ 4286} (\bibinfo{year}{1991}).

\bibitem{Hemstreet93p4238}
\bibinfo{author}{\bibfnamefont{L.~A.} \bibnamefont{Hemstreet}},
  \bibinfo{author}{\bibfnamefont{C.~Y.} \bibnamefont{Fong}}, \bibnamefont{and}
  \bibinfo{author}{\bibfnamefont{J.~S.} \bibnamefont{Nelson}},
  \bibinfo{journal}{Phys. Rev. B} \textbf{\bibinfo{volume}{{\bf 47}}},
  \bibinfo{pages}{\ 4238} (\bibinfo{year}{1993}).

\bibitem{Hybertsen86p2920}
\bibinfo{author}{\bibfnamefont{M.~S.} \bibnamefont{Hybertsen}}
  \bibnamefont{and} \bibinfo{author}{\bibfnamefont{S.~G.} \bibnamefont{Louie}},
  \bibinfo{journal}{Phys. Rev. B} \textbf{\bibinfo{volume}{{\bf 34}}},
  \bibinfo{pages}{\ 2920} (\bibinfo{year}{1986}).

\bibitem{Kleinman80p2630}
\bibinfo{author}{\bibfnamefont{L.}~\bibnamefont{Kleinman}},
  \bibinfo{journal}{Phys. Rev. B} \textbf{\bibinfo{volume}{{\bf 21}}},
  \bibinfo{pages}{\ 2630} (\bibinfo{year}{1980}).

\bibitem{Kleinman82p1425}
\bibinfo{author}{\bibfnamefont{L.}~\bibnamefont{Kleinman}} \bibnamefont{and}
  \bibinfo{author}{\bibfnamefont{D.~M.} \bibnamefont{Bylander}},
  \bibinfo{journal}{Phys. Rev. Lett.} \textbf{\bibinfo{volume}{{\bf 48}}},
  \bibinfo{pages}{\ 1425} (\bibinfo{year}{1982}).

\bibitem{Bachelet85p879}
\bibinfo{author}{\bibfnamefont{G.~B.} \bibnamefont{Bachelet}} \bibnamefont{and}
  \bibinfo{author}{\bibfnamefont{N.~E.} \bibnamefont{Christensen}},
  \bibinfo{journal}{Phys. Rev. B} \textbf{\bibinfo{volume}{{\bf 31}}},
  \bibinfo{pages}{\ 879} (\bibinfo{year}{1985}).

\bibitem{Perdew81p5048}
\bibinfo{author}{\bibfnamefont{J.~P.} \bibnamefont{Perdew}} \bibnamefont{and}
  \bibinfo{author}{\bibfnamefont{A.}~\bibnamefont{Zunger}},
  \bibinfo{journal}{Phys. Rev. B} \textbf{\bibinfo{volume}{{\bf 23}}},
  \bibinfo{pages}{\ 5048} (\bibinfo{year}{1981}).

\bibitem{GPL}
\bibinfo{author}{\bibnamefont{{Free Software Foundation}}},
  \urlprefix\url{http://www.gnu.org/copyleft/gpl.html}.

\end{thebibliography}

\end{document}